\documentclass[sn-mathphys-num]{sn-jnl}

\usepackage{graphicx}
\usepackage{multirow}
\usepackage{amsmath,amssymb,amsfonts}
\usepackage{amsthm}
\usepackage{mathrsfs}
\usepackage[title]{appendix}
\usepackage{xcolor}
\usepackage{textcomp}
\usepackage{manyfoot}
\usepackage{booktabs}
\usepackage{soul}
\usepackage{listings}
\usepackage[T1]{fontenc}
\usepackage{tabularx}

\theoremstyle{thmstyleone}

\theoremstyle{thmstyletwo}

\theoremstyle{thmstylethree}

\begin{document}
\newcommand{\mnras}{Mon. Not. R. Astron. Soc.}
\newcommand{\araa}{Annu. Rev. Astron. Astrophys.}
\newcommand{\apj}{Astrophys. J.}
\newcommand{\aj}{Astron. J.}
\newcommand{\apjl}{Astrophys. J.}
\newcommand{\apjs}{Astrophys. J.}
\newcommand{\aap}{Astron. Astrophys.}
\newcommand{\aapr}{Astron. Astrophys. Rev.}
\newcommand{\nat}{Nature}
\newcommand{\pasj}{Publ. Astron. Soc. Jpn}
\newcommand{\pra}{Phys. Rev. A}
\newcommand{\pasa}{Publ. Astron. Soc. Ast}

\title[Chemical Abundances near a Supermassive Black Hole]{Accurate Determination of Chemical Abundances near a Supermassive Black Hole}

\author[1]{\fnm{The XRISM} \sur{collaboration}}

\affil[1]{The list can be found on the last page}

\abstract{
The metal abundances in galactic nuclei carry key information on the
history of star formation and mass transfer in central regions of
galaxies. X-ray fluorescence analysis is a unique tool to reliably
measure the abundances of various elements via simple physics. Here we
present a new observation of the active nucleus in the Circinus Galaxy
with the XRISM satellite at unprecedented X-ray energy resolution. The
fluorescent iron-K$\alpha$ line profile modified by Compton scattering
indicates that the material responsible for its emission is cold,
metal-rich, and is located $\gtrsim$0.024 parsecs (pc) from the
supermassive black hole, consistent with the dusty torus region. The
abundance pattern derived from comparing fluorescent
  line intensities of different metals shows sub-solar ratios of
argon- and calcium-to-iron, and a super-solar ratio of nickel-to-iron.
This abundance pattern is best produced by a combination in number
fraction of $92^{+2}_{-4}$\% core-collapse supernovae from progenitor
stars less massive than $20^{+3}_{-2} M_\odot$ and $8^{+4}_{-2}$\%
type-Ia SNe. This suggests that gas feeding the
  super-massive black hole was enriched by recent core-collapse
  supernovae. Our findings imply that in metal-rich environments
stars more massive than about 20 $M_\odot$ directly
  collapse into black holes or make faint SNe without ejecting heavy
metals into the space.  }

\maketitle

The elemental abundance pattern is a crucial indicator for comprehending the history of metal production and supply in metal-enriched systems\cite{Maiolino19}.
It
reflects the history of past supernovae (SNe) of various types, roughly divided into Type~Ia SNe (SNe Ia) and core-collapse SNe (CCSNe).
Metals produced by SNe are eventually supplied to intracluster medium (ICM) and pollute its primordial gas synthesized by the Big Bang, consisting mainly of hydrogen and helium.
X-ray observations of hot ICM in galaxy
clusters \cite[e.g.,][]{Hitomi17} suggest that the
metal abundance ratios integrated over the cosmic time
are close to the solar values, 
which refer to the protosolar table of \citet{Lodders09} in all X-ray spectral models employed in this work. This abundance pattern
can be
explained by a combination of about 20\% of SN Ia and about 80\% of
CCSN events according to current SN nucleosynthesis
models\cite[][]{Nomoto13,Fink14}.
However,
probing the chemical abundance in the very central region of a galaxy,
which holds a key to deciphering its evolutionary history,
has remained challenging.
In the UV, optical, and
near-infrared bands,
metal abundances in galaxies hosting super-massive black holes (SMBHs) are measured with emission-line ratios from
nebulae via photoionization modelling, which traces primarily the interaction of optical and UV photons with the surrounding gas
\cite[e.g.,][]{Nagao06a, Nagao06b, Floris24}.
However, the results sensitively
depend on the assumed density distribution, the shape of the ionizing
continuum, and the degree of dust
depletion \cite[][]{Nagao06a,Nagao06b},
leaving large systematic uncertainties.
By contrast, 
the interaction
between X-rays and matter is simpler, allowing for a nearly unbiased
probe of all matter including gas and dust. 
A fluorescence line is emitted after photoelectric absorption of a high energy photon by an atom, and hence its intensity represents the amount of each metal contained in X-ray irradiated medium. 
Thus, diagnostics based on fluorescence lines in
the X-ray spectrum reflected off matter surrounding the SMBH
enable a more direct estimate of
the elemental abundances in the environments of Active Galactic Nuclei (AGNs).
Thanks to its unprecedented energy resolution,
the Resolve spectrometer onboard the recently launched X-ray observatory XRISM\cite{Tashiro25}
enables the
precise measurement of faint fluorescence
emission lines from elements that were previously beyond reach.

In this study, we focus on the Circinus Galaxy (hereafter Circinus), a
nearby (4.2 Mpc; \cite{Freeman77}), 
spiral galaxy hosting the nearest Seyfert 2 nucleus \cite{Oliva94}.
Previous X-ray
observations have revealed that Circinus harbours a Compton-thick AGN
\cite[e.g.,][]{Matt96, Arevalo14, Tanimoto19},
where the central engine is heavily obscured by 
intervening material with a hydrogen column density of $N_{\rm H} > 10^{24}$ cm$^{-2}$. This makes it
an ideal target for studying the properties of the obscuring torus and
its reflection spectrum.

\begin{figure}[ht]
\centering
\includegraphics[width=\linewidth]{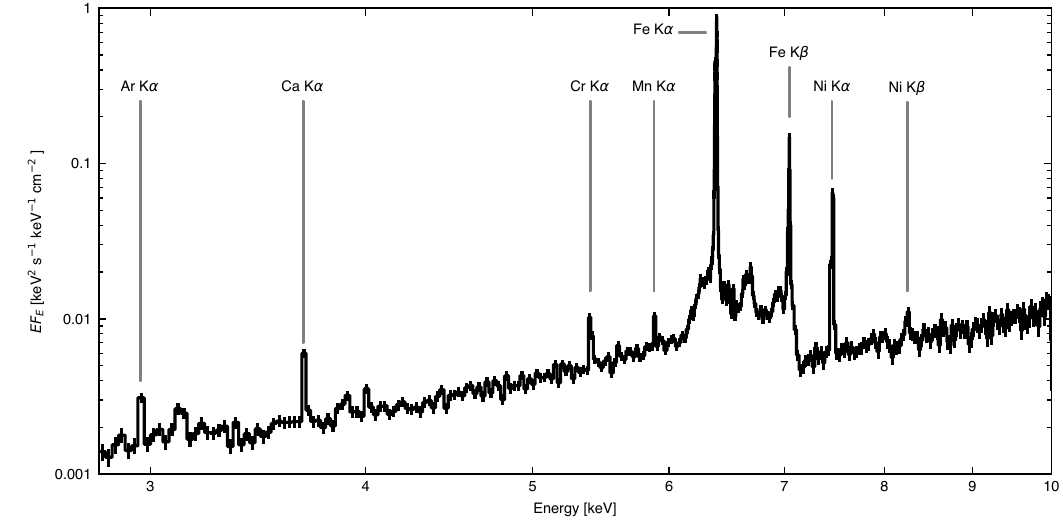}
\caption{{\bf XRISM/Resolve spectrum of the Circinus galaxy in the
  2.8--10 keV band.} It is folded by the instrumental response in units of $E F_E$, where $F_E$ is the energy flux at energy $E$ (i.e., corrected for the effective area). 
  The spectrum is binned according to the ``optimal binning'' technique\cite{Kaastra16}, which is used for the spectral analysis throughout the paper.
  The non X-ray background component is subtracted. 
  Major fluorescence lines from cold
  matter are labelled. 
The error bars denote the 1$\sigma$ confidence limits in photon statistics.
}\label{fig:spec}
\end{figure}

\begin{figure}[ht]
\centering
\includegraphics[width=\linewidth]{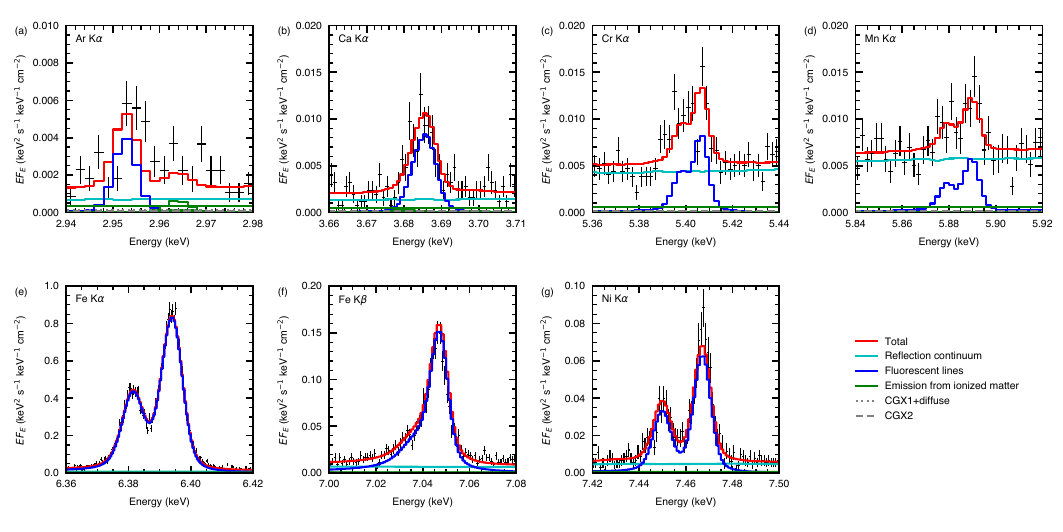}
\caption{{\bf Enlarged plots of the XRISM/Resolve
      spectrum of the Circinus galaxy around the major fluorescence lines
      from neutral matter, overlaid with the best-fit model.} (a) Ar K$\alpha$. (b) Ca K$\alpha$. (c) Cr K$\alpha$.
      (d) Mn K$\alpha$. (e) Fe K$\alpha$. (f) Fe K$\beta$. (g) Ni K$\alpha$.
    They are folded by the instrumental response
in units of $E F_E$, where $F_E$ is the energy flux at energy $E$. 
The observed data are plotted in black with error bars denoting the 1$\sigma$ confidence limits in photon statistics.
The blue, cyan, 
  green, and red lines represent the emission lines from cold matter, the reflection continuum from cold matter, emission 
  from ionized matter together with the Thomson-scattered component, and the total, 
  respectively, based on our best-fitting model (Methods Section~\ref{sec:specmodel}).
  For visualization, the bin size is set at either 2 eV (Ar K$\alpha$, Cr K$\alpha$, and Mn
  K$\alpha$), 1 eV (Ca K$\alpha$, Fe K$\beta$, and Ni K$\alpha$), or 0.5 eV (Fe K$\alpha$) in these plots.
  }\label{fig:specline}
\end{figure}

Circinus was observed by XRISM
on 2024 February 8--12, with a total
exposure time of 309 ksec.  Nearly simultaneous NuSTAR and XMM-Newton
observations were carried out to extend our spectral coverage
to higher energies and to obtain better spatial resolution below 10 keV.
The
overall XRISM/Resolve spectrum
covering an approximately $3' \times 3'$ area centred at the Circinus nucleus is plotted in Fig.~\ref{fig:spec}, where a wealth of emission lines from various elements are
  noticeable. Its enlarged view around the fluorescence lines of
  argon, calcium, chromium, manganese, iron, and nickel is shown in
  Fig.~\ref{fig:specline}(a)-(g).

Our primary goal is to model the reflection spectrum from the AGN
torus\cite{RamosAlmeida17}, most likely located at
about $0.1-10$ pc from the SMBH \cite{Wada16} in Circinus, with a
particular focus on the fluorescent lines from cold matter. To this
end, we employed an updated version of the state-of-art clumpy torus
model, XCLUMPY \cite{Tanimoto19}, which accounts for the non-uniform
distribution of matter in the AGN environment (Methods Section~\ref{sec:specmodel}). To accurately
model the fluorescent line shapes, we incorporated the natural line
widths determined from laboratory measurements. This approach is
crucial for taking full advantage of the high spectral resolution
provided by XRISM/Resolve.  Considering the limited photon
statistics below $\approx$2.5 keV in the Resolve spectrum (see Extended Data
Fig.~1), here we focus on the lines from Ar and heavier
elements. Accordingly, the abundances of Ar, Ca, Cr, Mn, Fe, Ni
were left as free parameters.

We simultaneously fit the XRISM/Resolve spectrum in the 2.0--10 keV
band and the NuSTAR one in the 8--70 keV band. The model consists of
the torus reflection component described above, transmitted component,
emission from ionized gas,
and emission from
contaminating sources other than the AGN (Methods Section~\ref{sec:contamination}). To take into account Doppler
broadening of lines, we convolve the line components from the torus
with a Gaussian kernel.
This model is found to well reproduce the whole spectra.

\begin{figure}[ht]
\centering
\includegraphics[width=\linewidth]{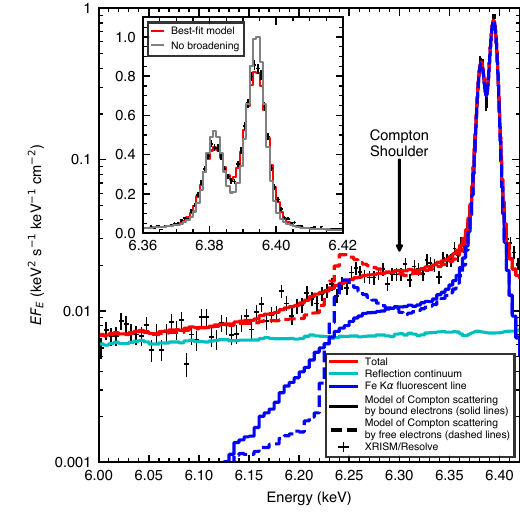}
\caption{{\bf A zoom-in picture of XRISM/Resolve
      spectrum of the Circinus galaxy around the Fe K$_{\alpha}$ emission line.}
    They are folded by the instrumental response
in units of $E F_E$, where $F_E$ is the energy flux at energy $E$. 
    The observed data are plotted in black with error bars denoting the 1$\sigma$ confidence limits in photon statistics.
  The blue, cyan, and red lines represent the models of the emission lines from cold
matter, the reflection continuum from cold matter, and the total,
respectively. The dashed lines denote the case where
  Compton scattering from free electrons at zero temperature is
  considered.  In the inset, the observational
data are compared against our best-fit model (red) and a model to
which no Gaussian broadening is applied (grey). This shows that
Doppler broadening is very small compared with the intrinsic width of
the Fe $K_{\alpha}$ line.
}
\label{fig:cs}
\end{figure}

We have obtained the best quality spectrum of a
narrow iron K$\alpha$ line (Fig.~\ref{fig:cs})
ever observed from an AGN (comparison with the Chandra/HETG data is
shown in Extended data Fig.~2).  It is characterized by
its narrowness, with Full Width Half Maximum
  (FWHM) of $210\pm10$ km s$^{-1}$,
which was overestimated in previous studies
\cite[e.g.,][]{Shu11,Andonie22},
most probably due to the
non-consideration of the fine structure and/or natural broadening in the intrinsic line profile\cite{Hoelzer97} and to
spatial-spectral degeneracies in grating data\cite{Uematsu21}.
The width is larger than those of atomic and molecular lines
at 0.5--4 pc ($115.2\pm0.8$ km s$^{-1}$ in FWHM for [C I]$^3$P$_1$--$^3$P$_0$
and 141.1$\pm$0.8 km s$^{-1}$ in FWHM for CO($J=3\rightarrow2$) where $J$ is the total angular momentum quantum number), but is smaller than that
of ionized gas in the central core ($<0.5$ pc), 
traced by the atomic hydrogen recombination line, 
H36$\alpha$, at 135 GHz with ALMA ($393\pm44$ km s$^{-1}$ in FWHM) \cite{Tristram22, Izumi23}.
To infer the innermost radius of the line emitter, 
we convolve the intrinsic line profile with 
a kernel that represents Doppler broadening by a Keplerian disk 
(\texttt{rdblur} in XSPEC) instead of a Gaussian.
  Adopting an inclination of 79.1 degree, the best-fit value in the XCLUMPY spectral fitting (Extended Data Table~1),
and assuming an outermost radius of $10^8 r_{\rm g}$
($r_{\rm g} \equiv GM/c^2$ is the
gravitation radius, where $G$, $M$, and $c$ are the gravitational constant, black hole mass, and light velocity, respectively),
which roughly corresponds to the size of the nuclear dust continuum emission imaged by ALMA \cite{Tristram22},
we obtain an emissivity index of $q=-1.7\pm0.1$ and an innermost radius of
$(3.0\pm0.6) \times 10^5$ $r_{\rm g}$,
which corresponds to $0.024\pm0.007$ pc
for an SMBH mass of $(1.7\pm0.3)\times10^6 M_{\odot}$ \cite{Greenhill03}.
This is almost consistent with the dust
sublimation radius of the torus in Circinus, 0.05$\pm0.02$ pc
  \cite{Uematsu21}, assuming a factor of 2 uncertainty in the bolometric luminosity.
Another key feature is the smeared shape of the
low energy side of the Compton shoulder structure
\cite[see][]{Odaka16,Dimopoulos24}
at 6.24 keV (corresponding to the backscattering case by a free
electron at rest).
When X-rays are scattered by electrons bound in atoms, the edge
feature is blurred by an order of the electron binding energies
due to the Doppler broadening.
Hence, our result 
suggests that
the scatterer is likely to be cold (not-ionized) material. Within our
measurement uncertainties, the centroid energy of the Fe K$\alpha$
line is consistent with the rest-frame energy of cold atom (Methods Section~\ref{sec:specmodel}),
indicating that the bulk of the reflecting material is not undergoing
significant systematic motion relative to the host galaxy, unlike the
optical lines from the ionization cone that show blueshifts of
$\gtrsim 200$ km s$^{-1}$ \cite{Fischer13,Kakkad23}.  All these
considerations
suggest that
the Fe K lines
predominantly come from an
innermost part of the ``cold'', dusty torus,
which dominates the mass distribution of the surrounding matter. 
We note, however, that
although our Circinus spectrum does not require 
contributions
from 
the broad line region as discussed by \citet{Gandhi15} and \citet{Andonie22}, it may be due to the heavy obscuration by the torus that completely blocks emission inside it. In fact, XRISM observations of less obscured AGNs commonly show broader components in Fe K$\alpha$ \cite[e.g.,][]{XRISM2024NGC4151}.

We have accurately
determined the abundance ratio of iron to hydrogen is $2.3\pm0.1$
times the solar value
through a joint analysis of the broadband spectra and
iron K$\alpha$ profile (see Method~\ref{sec:specmodel} for details).
The shape of the broadband reflection continuum basically
  determines the geometry and hydrogen column density of the torus. Then, as in previous studies \cite{Matt97}, the
  equivalent width of iron K$\alpha$ can be used to infer the Fe/H
  abundance ratio. 
  To obtain a tighter constraint on Fe/H, we additionally utilize the intensity of iron K$\alpha$ and its
  Compton-shoulder fraction $f_{\rm CS}$, which are roughly
  proportional to the amount of iron atoms and hydrogen atoms in the
  reflector, respectively\cite{Odaka16}.
  We have confirmed that our results on the Fe/H abundance ratio is robust
  against different parameters of the torus geometry (Methods Section~\ref{sec:torusgeometry}).

\begin{figure}
\centering
\includegraphics[width=\columnwidth]{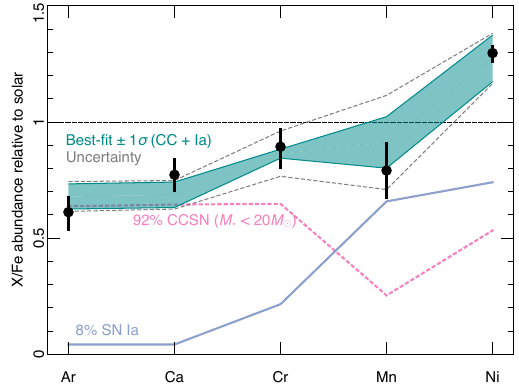}
\caption{
{\bf 
Observed elemental abundances of the Circinus centre and theoretically calculated values with an upper mass limit of 20 $M_\odot$for CCSN progenitors.} Relative abundances with respect to iron determined with XRISM (X/Fe, filled circles) normalized to the corresponding proto-solar values (horizontal dashed line). The error bars denote the 1$\sigma$ confidence limits on the measurement, estimated by Markov Chain Monte Carlo method.
  The green shaded area represents the 1$\sigma$ statistical uncertainty region of the best-fit model ($\sigma_\textup{stat}$)
  consisting of 92\% CCSNe from 
  progenitor stars with zero-age main-sequence masses ($M_*$) smaller than $20 M_\odot$
  and 8\% near-$M_\textup{Ch}$
  SNe~Ia in number fraction.
The contribution that CCSNe and SNe~Ia make to the 
X/Fe are represented by the magenta and blue curves, respectively. Here we adopt the nucleosynthesis models for CCSNe \cite{Nomoto13} and SNe~Ia \cite{Fink14, Shen18}.
Thin grey-dashed lines show the total uncertainty, including differences in the predicted yields
by alternative SN~Ia and CCSN modelling
as $\sqrt{\sigma_\textup{stat}^2 + \sigma_\textup{Ia}^2 + \sigma_\textup{CC}^2}$ (see Methods Section~\ref{sec:snmodels}).
}\label{fig:cut}
\end{figure}

Fig.~\ref{fig:cut} presents the observed abundance ratios relative
to iron in the Circinus centre.
The Ar/Fe and Ca/Fe ratios 
exhibit sub-solar (0.6--0.8) values,
whereas the Ni/Fe ratio is super-solar (1.3).
Our abundance pattern 
differs from that of the Perseus cluster core\cite{Hitomi17, Simionescu19}, which follows the solar abundance ratios including Ni/Fe, and 
from nuclear star clusters in the Milky Way\cite{Nandakumar25}, which
show sub-solar Ca/Fe but super-solar Mn/Fe ratios.

To interpret our result, 
the observed abundance pattern is modelled by 
using a linear combination of CCSN
and SN~Ia nucleosynthesis predictions \cite[e.g.,][]{Simionescu19},
yielding the abundance ratio of element X to iron as
\begin{eqnarray*}
\frac{A_\textup{X}}{A_\textup{Fe}} = \frac{r^\textup{Ia} A_\textup{X}^\textup{Ia} + A_\textup{X}^\textup{CC}}{r^\textup{Ia} A_\textup{Fe}^\textup{Ia} + A_\textup{Fe}^\textup{CC}}\textup{,}
\end{eqnarray*}
where $r^\textup{Ia}$ denotes the number ratio of SNe~Ia to CCSNe.
Progenitors for each SN type are assumed to have the solar metallicity ($Z = 0.02$).
The Salpeter initial-mass function (IMF, index $-2.35$, \cite{Salpeter55}) is employed to integrate
yields over CCSNe progenitor masses ranging from $9$--$M_\textup{CCSN, up}$;
\begin{eqnarray*}
  A_\textup{X}^\textup{CC} = \frac{\int_{9 M_\odot}^{M_\textup{CCSN, up}} M^{-2.35} A_\textup{X}(M) dM}{\int_{9 M_\odot}^{M_\textup{CCSN, up}} M^{-2.35} dM}\textup{,}
\end{eqnarray*}
where $A_\textup{X}(M)$ represents the yield of element X as a function of the progenitor mass $M$.
When $A_\textup{X}(M)$ is available only at discrete values of $M$ in the literature,
we simply refer to the value closest to each calculation point in the integration.
Here we assume that all stars 
formed according to the IMF in the integrated mass range have already undergone CCSNe.
This assumption is valid because the young nuclear stellar population in Circinus
has estimated ages of $4\times10^7-1.5\times10^8$ yr \cite{Maiolino98},
which is comparable or longer 
than the life time of a star with an initial mass of 9 $M_\odot$.
For the metal yields, 
we refer to the well-established CCSN model by \citet{Nomoto13},
alongside with the SN~Ia yields from near-Chandrasekhar-mass
($M_\textup{Ch}$) progenitors \cite[][]{Fink14} and merging
sub-$M_\textup{Ch}$ ones \cite[][]{Shen18}.
Note that CCSNe
from high mass ($\gtrsim 20 M_\odot$) progenitors produce a large
amount of $\alpha$ elements relative to iron, compared with SNe~Ia
\cite[][]{Nomoto13, Fink14}.

While detailed studies on the origin of the metals in Milky Way
stars\cite[e.g.,][]{Weinberg19} consider 
star formation history (SFH) to explain the evolution of chemical
elements over the cosmic time, in this paper we focus on
the
relative contributions from different explosion sites (e.g., the
initial-mass dependent SN yields), which determine the abundance
ratios among the heavy elements.
We do not assume that the SNe~Ia and CCSNe originate from the same
stellar population,
or consider SFHs representing the detailed metal enrichment history.
The delay time distribution of SNe~Ia after birth
typically ranges 100 to 1000\,Myr \cite[e.g.,][]{Chen21},
whereas that of CCSNe is much shorter ($<50$\,Myr for a single massive progenitor, \cite{Zapartas17}).
Thus, the fraction of CCSNe estimated by our analysis provides
key information on contribution from young star formation
at the central region of Circinus.
Throughout the paper, we assume that the AGN does not influence
stellar evolution and distribution of elements.

As a reference model, we first consider a wide mass range for CCSN
progenitors by setting $M_\textup{CCSN, up} = 40 M_\odot$
with contribution of near- and sub-$M_\textup{Ch}$ SNe~Ia.
The five abundance ratios (Ar/Fe, Ca/Fe, Cr/Fe, Mn/Fe, and Ni/Fe) are
compared with model predictions by varying the number fractions of
three SN types (i.e., with two independent free parameters). The fitting is performed by
minimizing the $\chi^2$ statistics. We find that
the observed pattern can be roughly explained by a combination of
60\%
CCSNe and 40\% SNe~Ia in number fraction
(Table~\ref{tab:frac};
Extended Data Fig.~3).
However, the fitting is not statistically acceptable 
with reduced $\chi^2$ of 3.3. Furthermore,
the typical SN~Ia
fraction even in ancient systems, such as galaxy clusters and
early-type galaxies, varies from 10\% to 25\% \cite[e.g.,][for a recent
  comprehensive study]{Simionescu19};
therefore, the
substantial SN~Ia contribution higher than this level
is hard to interpret for the Circinus
centre, wherein a continuous star formation is
identified\cite{Marconi94,Mueller-Sanchez06}.
Thus, this reference model is not favoured due to both statistical
and physical issues.

Alternatively, we select a different mass range for CCSN progenitors
by employing IMF-weighted integrations with the upper mass limits
in the range of $M_\textup{CCSN, up} =$\,18--35\,$M_\odot$.
As summarized in Table~\ref{tab:frac}, the models with $M_\textup{CCSN, up} < 30\,M_\odot$
give statistically much better fits to the observed abundance pattern than
the SNe~Ia dominant model,
and produce a natural enrichment scenario, with a dominant contribution from CCSNe.
Fig.~\ref{fig:cut} shows the best-fitting combination model,
where the number fraction of CCSNe with progenitors less massive than $20^{+3}_{-2} M_\odot$
is $92^{+2}_{-4}$\%, globally reproduces the abundance ratios of 
Ar, Ca, Cr, Mg, Ni, relative to Fe.
The attached errors are statistical ones at $1\sigma$ confidence
limits; we estimate that in $M_\textup{CCSN, up}$ from the
$M_\textup{CCSN, up}$ versus $\chi^2$ plot with 1 $M_\odot$ steps.
Methods Section~\ref{sec:snmodels} summarizes
other modelling trials
assuming different SN Ia models or CCSNe models with alternative yield calculation, IMF, hypernova contribution, and progenitor metallicity.
We confirm that our conclusions are not
  affected. The systematic uncertainty in the CCSN or SN~Ia fraction
is found to be 2\%, which is smaller than the statistical one.

Our findings indicate two important consequences on SMBH-galaxy co-evolution and on the mechanisms of CCSNe.
The SN composition suggests that the SMBH is accreting gas whose metal is
largely enriched by recent CCSNe rather than old gas produced by SNe~Ia.
This does not contradict the ``inside-out'' galaxy formation
scenario, where the core of a galaxy formed in the very early universe
\cite{Baker25}, if gas in the torus region $<$10--20 pc is continuously
supplied from the outer disk, replacing old gas (Methods Section~\ref{sec:masstransfer}).
The co-existence of nuclear starburst and AGN activities 
is in line with models where SNe play a role for 
obscuration of the central engine \cite{Wada12}.
Also, high metallicity in galactic nuclei implies the importance of AGN feedback as a metal enrichment process on galaxy scales and beyond\cite{VillarMartin24}.

The presence of an upper mass limit for CCSN progenitors
at the galactic central region has a critical
implication for our understanding of the fate of massive stars.
Our scenario is consistent with the suggestion by \citet[][]{Mao21} based on
the chemical abundance measurement of the hot atmosphere of a starburst galaxy.
Some
theories postulate that stars with zero-age main-sequence masses ($M_{\rm ZAMS}$)
of a few tens of solar masses will not undergo CCSN explosions but
directly collapse into black holes (BHs) \cite[e.g.,][]{Heger03} or make faint SNe without emitting heavy metals \cite{Lovegrove13}.
Observationally, while there are
a few possible examples for such BH-forming events in the local
universe\cite{Adams17},
it is still controversial whether this channel
can make a substantial contribution to the BH population.
Our finding supports the direct BH-forming collapses or faint SNe
for stars with $M_{\rm ZAMS} > 20^{+3}_{-2} M_\odot$ in metal-rich environments
unless massive stars are not formed due to, for example, a severely distorted IMF.
This scenario may also solve the ``red supergiant
(RSG) problem'' \cite[e.g.,][]{Smartt09,Smartt15} that no CCSNe from RSG stars
with $M_{\rm ZAMS} > 18 M_\odot$ have been found in the local universe
despite the existence of more massive RSGs.
In addition, it gives an
important clue to understand nucleosynthesis history in the
universe. We infer that, in the early universe when the gas is metal
poor, very massive stars (e.g., $M_{\rm ZAMS}$ of several tens of
solar masses beyond the BH-forming mass window) explode as CCSNe and
produce a large amount of alpha elements, in order to explain the
solar abundance ratios in galaxy clusters \cite{Hitomi17, Simionescu19}.

\clearpage

\flushleft{{\Large\bf Methods}}
  
\section{XRISM Data Reduction}\label{sec:xrismreduction}

We used Heasoft version 6.34 and XRISM Calibration Database version
20240815 for the data reduction. The Resolve data were reduced
according to the XRISM Quick Start Guide
v2.3
and additional screening was applied using the standard
energy-dependent rise time cut \cite{Mochizuki24}. The
response matrix file (RMF) was generated using \textsc{rslmkrmf}, and
its normalization was calculated based on the
event grade distribution
in the cleaned event file. The cleaned file initially included
low-resolution secondary (Ls) events, which were not caused by the
source but resulted from the secondary-pulse detection algorithm
processing clipped pulses. Since almost all events should have a
high-resolution primary
grade (Hp) due to the count rate of $10^{-3}$–$10^{-1}$ s$^{-1}$ per pixel,
the Ls events were removed from the cleaned event file before
calculating the RMF.  The ancillary response file (ARF) was created
with \textsc{xaarfgen}, assuming a point-like source at the aim point
as input.
Only Hp events were used for spectral
analysis.

The gain and thus the energy scale of Resolve is sensitive to its
environment. Thus, the gain is monitored using onboard calibration
sources during fixed fiducial intervals in the standard Resolve
calibration strategy \cite{Porter24}. The fiducial intervals are then
interpolated to reconstruct the time dependent energy scale. The
efficacy of this strategy is monitored by a continuously illuminated
calibration pixel that is located just outside the aperture of the
instrument, but part of the 50 mK detector array. The calibration pixel
is continuously illuminated by a heavily collimated $^{55}$Fe source but is
corrected using the same fiducial intervals as the main array. For
this observation, the calibration pixel energy scale offset was
measured to be 0.07 eV at 6 keV. This is added in quadrature to the
systematic uncertainty in the Resolve energy scale which is the
standard method for Resolve observations. The systematic uncertainty
in the Resolve energy scale is 0.3 eV from 5.4--9 keV (the range of the
suite of on-board calibration sources)\cite{Eckart24}. Added in
quadrature, the systematic uncertainty in the energy scale for this
observation is thus 0.31 eV. During the fiducial intervals, the energy
resolution of the composite spectrum of the main array was 4.44 eV at
5.9 keV which is in-line with the standard XRISM/Resolve RMF.

The XRISM/Resolve main array is composed of 36 pixels, one of which is
the calibration pixel which is outside of the instrument aperture. For
this observation, 34 main array pixels were used in the
analysis. Pixel 27 was not used because it has been shown to exhibit
gain jumps which are not correctable with the sparse gain fiducials
used in the standard energy scale reconstruction method.

The Non-X-ray Background (NXB) event file was generated using the
\textsc{rslnxbgen} task, and it was screened with the same criteria as
the source data, excluding Pixel 27 and retaining only Hp events. The
NXB spectrum was fitted using a model provided by the XRISM
Calibration team, which includes a power-law component and 17 narrow
emission lines modeled by Gaussian profiles. These lines correspond to
Al K$\alpha$1/K$\alpha$2, Au M$\alpha$1, Cr K$\alpha$1/K$\alpha$2, Mn
K$\alpha$1/K$\alpha$2, Fe K$\alpha$1/K$\alpha$2, Ni
K$\alpha$1/K$\alpha$2, Cu K$\alpha$1/K$\alpha$2, Au
L$\alpha$1/L$\alpha$2, and Au L$\beta$1/L$\beta$2. Extended Data Fig.~1
plots the observed spectrum of Circinus (not corrected for effective area)
and this NXB model.

\section{NuSTAR Observations and Data Reduction}\label{sec:nustarreduction}

The hard band spectrum of the Circinus galaxy was obtained with NuSTAR
on February 2 2024, with a total exposure time of 24 ksec (ID
60901013002). We reprocessed the data from the two FPM detectors,
following the standard procedure.
The source spectrum 
was extracted from a circular region with a radius of 60 arcsec
centred on the source peak. The background was
taken from a circular region with a radius of 60 arcsec that is not
contaminated by bright Chandra-detected sources or the PSF wings of
the nucleus. The source spectra were binned to contain at least 50
counts per bin.

\section{Analysis of Contaminating Sources}\label{sec:contamination}
\label{sect:cont}

The Resolve spectrum of the Circinus galaxy is mainly contaminated by
three sources: CGX1 (ultraluminous X-ray source), CGX2 (supernova remnant), and diffuse emission (see \cite{Smith01,Marinucci13,Arevalo14,Kawamuro19}).
To constrain the impact of these contaminating sources,
we utilized the high-spatial-resolution image obtained with
XMM-Newton. The Circinus galaxy was observed with XMM-Newton from
February 2 2024 to February 3 2024, with a total exposure time of 115
ksec (ID 0932990101). The unfiltered data was reprocessed following
the standard procedure with \textsc{sas v21.0.0}. We extracted
two types of 
spectra with the following recipe. First, the ``XMM-all''
spectrum, which contains nearly all of the contaminating sources in
the Resolve spectrum, was extracted from a circular region with a
radius of 90 arcsec centred on the Circinus galaxy, 
thus commensurate to the Resolve field-of-view.
Here we utilized the pn data, which have better photon statistics than
the MOS1 or MOS2 data.
The background
spectrum for the ``XMM-all'' spectrum was extracted from a
source-free annulus with a inner radius of 100 arcsec and a outer
radius of 140 arcsec. 
Second, the ``XMM-CGX2'' spectrum, which represents the
source spectrum of CGX2, was extracted from a circular region with a
radius of 8 arcsec centred on CGX2.
For this we utilized the MOS1 data, which have the best 
spatial resolution among the 3 EPIC cameras.
The background spectrum for the
``XMM-CGX2'' spectrum was extracted from three circular regions,
each with a radius of 8 arcsec and located at the same distance from
the Circinus galaxy.
The image of MOS1
is displayed in Supplementary Fig.~1, where the source and background
regions for the ``XMM-CGX2'' spectra are indicated, together with
the corresponding source region for the ``XMM-all'' spectrum. 
All the source spectra are binned to contain at
least 50 counts per bin. Note that there is a known issue in the cross
calibration between NuSTAR and XMM-Newton \cite{fuerst22}. To correct
this issue, we always applied \texttt{applyabsfluxcorr=yes} option in
\textsc{arfgen}.

To better constrain the CGX2 spectrum, we also utilized the high
spatial and energy resolution spectra obtained with Chandra/HETG. The
Circinus galaxy was observed with Chandra/HETG for several times from
December 8 2008 to March 4 2009 (ID 10223, 10224, 10225, 10226, 10832,
10833, 10842, 10843, 10844, 10850, 10872, 10873). We reprocessed all
the spectra following the standard procedure with \textsc{ciao
  v4.17}. The combined spectra were binned to contain at least one
count per bin.

First, we calibrated the spectral model for CGX2 utilizing the Chandra/HETG spectra. We simultaneously fitted the HEG (first order), the MEG (second order), and the zeroth order spectra with a constant temperature plane-parallel shock plasma model modified by Galactic absorption of $N_\mathrm{H}=7.02\times 10^{21}\,\mathrm{cm}^{-2}$ \cite{Willingale13}. The model is
expressed as follows in \textsc{XSPEC} terminology:
\begin{equation*}
	\mathrm{CGX2} = \texttt{TBabs} * \texttt{bvpshock}
\end{equation*}
We successfully reproduced the spectra (C-stat $=4454.6$ for 4536 d.o.f.)
with this model, which gives a similar spectrum to that adopted by \citet{Arevalo14}.
Note that the \texttt{TBabs} model implements the photoelectric absorption cross section in the interstellar medium given by \citet{Wilms00} and assumes one solar abundances.

Supplementary Fig.~2 plots the Chandra/HETG spectra together with the best-fit model.
Next, we fit the ``XMM-CGX2'' spectrum with the
calibrated model described above. Here we left
the normalization to vary
to treat the known time variability of CGX2
\cite{Quirola-Vasquez19}. We well reproduced the ``XMM-CGX2''
spectra with $\chi^2=21.1$ for 26 d.o.f..
The combined spectrum of CGX1 and the diffuse emission was approximated by a single power-law with a cutoff at 10 keV, which was expressed as \texttt{zcutoffpl} in \textsc{XSPEC} terminology.
Although \citet{Arevalo14} adopted a simple absorbed power-law model for CGX1, the difference little affects the spectral fitting of the nuclear emission, which is much harder than the CGX1 spectrum.

The model to fully characterise the contribution of the contaminating sources in the
XRISM/Resolve and NuSTAR apertures was determined by fitting them together with the XMM-Newton/EPIC-pn
spectra, by using models representing independently the AGN, CGX1+diffuse and CGX2 contributions.
The spectral model for the AGN component is described in Section~\ref{sec:specmodel}. The model and the parameters of CGX2 were fixed at the values as derived from the XMM-CGX2 spectra. We included a cross-normalization factor (relative to Resolve) to correct for a possible difference in the absolute flux calibration among different instruments.
Note that the normalization of CGX2 was set to keep consistency with the cross-calibration factor.

\section{Spectral Model Description}\label{sec:specmodel}

As mentioned in the main text, the reflection component from an AGN
torus carry critical information on its geometry, hydrogen column
density, and metal abundances. Basically, the broad continuum is
determined by the torus geometry and hydrogen column density
\cite{Ikeda09,Paltani17,Balokovic18,Tanimoto19,Buchner19}, whereas the
equivalent width \cite{Matt97} and Compton shoulder fraction
\cite{Odaka16} of iron K$\alpha$ are determined by the Fe/H abundance
ratio. It is noteworthy that the estimated torus parameters also
depend on the intrinsic spectrum (photon index and high energy
cutoff); for instance, a harder spectrum produces a stronger iron
K$\alpha$ line because of a larger number of photoionizing photons
above the iron K-edge energy.  It is not easy to accurately
determine the intrinsic spectrum in a heavily Compton thick AGN where
the transmitted component is not directly observable.
Thus, it is crucial to simultaneously fit the broadband continuum spectrum
and high energy-resolution spectrum covering the iron K$\alpha$ band to
best constrain the torus parameters by solving the degeneracy among them.

To derive the physical properties of the Circinus galaxy, we developed a new model that is applicable to the state-of-art high-resolution spectrum of XRISM. The basic idea of this model is based on the XCLUMPY model \cite{Tanimoto19}, which was designed to reproduce the X-ray reflection from a clumpy torus based on the same formalism as in the CLUMPY model \cite{Nenkova08b}.
In XCLUMPY, spheres with a uniform density are distributed
  according to a Gaussian distribution (with a standard deviation of $\sigma$)
  in the elevation direction and a power law distribution with an index of $-0.5$ in the radial direction.
The inner and outer radii of the torus are set at $r_{\rm in}=0.05$ pc and $r_{\rm in}=1.0$ pc, respectively, and the radius of each clump is set at $R_{\rm clump}=$0.002 pc; note that XCLUMPY is a scale-free model (i.e., the output spectra are identical as far as the ratios among these three parameters are kept the same).
  The expected value of number of blobs along the sight line on the equatorial plane is fixed at 10. The free parameters regarding the torus properties
are the torus angular width ($\sigma$), the total hydrogen column density in the equatorial plane ($N_{\rm H}^{\rm Equ}$), and the inclination angle ($i$).
A cross section view of the distribution of clumps in XCLUMPY is shown in Supplementary Fig.~3(a).
We expanded the parameter ranges to treat variable abundances and implemented the natural widths of prominent fluorescent lines referring to the following databases: \citet{Hoelzer97} (Fe K$\alpha$, Fe K$\beta$, Co K$\alpha$, Co K$\beta$, Ni K$\alpha$, Ni K$\beta$, Cr K$\alpha$, Cr K$\beta$), Hitomi Calibration report (Mn K$\alpha$, Mn K$\beta$), \citet{Ito16} (Ca K$\alpha$, Ca K$\beta$), and \citet{Krause79} (Ar K$\alpha$).

We fit the Resolve and NuSTAR spectra with the updated version of XCLUMPY.
The broadband Resolve-NuSTAR model is attributed to contributions from
different components, as follows:
\begin{equation*}
  const * (\mathrm{AGN} + \mathrm{CGX2} +  \mathrm{CGX1+diffuse})
\end{equation*}
The first term (\texttt{const}) is a cross-normalization factor (relative to Resolve) to correct for a possible difference in the absolute flux calibration among different instruments. The model and the parameters of CGX2 and CGX1+diffuse were fixed
to the values found in Section~\ref{sect:cont}.
Their summed contributions in the XRISM/Resolve spectrum are
approximately 1.1\% at 5 keV and 0.2\% at 10 keV.

In \textsc{XSPEC} terminology, the AGN component is expressed as follows:
\begin{eqnarray*}
  \mathrm{AGN} &=& \texttt{TBabs} * \texttt{TBvarabs} * (\texttt{TBvarabs}*\texttt{cabs}*\texttt{zcutoffpl}\\
  && + \texttt{xclumpy$_\mathrm{R}$.fits} + \texttt{gsmooth} * \texttt{xclumpy$_\mathrm{L}$.fits} + 3\times\texttt{pion} + \texttt{zcutoffpl})
  \end{eqnarray*}
The first term (\texttt{TBabs}) represents the Galactic absorption
and the second one (\texttt{TBvarabs}) takes into account foreground absorption towards the nucleus
in the host galaxy \cite{Roche06,Tristram07} with the same chemical abundances as in the XCLUMPY model.
In the parenthesis, the first combination of terms describes the direct
component transmitted through the torus.
The second and third
sets of terms represent the reflection continuum and
the fluorescence lines from the torus. To detect energy shifts from
the literature values \cite{Hoelzer97}, the redshift parameters are
allowed to vary. The line components are convolved with a Gaussian
kernel to determine additional broadening. 
As discussed in the main paper,
  the bulk of the iron emitting material is constrained to a location at about 0.03~pc. 
Based on our analysis of the Chandra imaging data, 
spatially extended Fe K$\alpha$ line emission on
a 100 pc scale, revealed by \citet{Marinucci13},
makes only a minor ($<1\%$) contribution to the XRISM
Fe K$\alpha$ signal.
The metal abundances, photon index, cutoff energy, and
normalization are linked to those of the direct component.
The last set of
terms describe the emission from photoionized plasmas
seen in the Resolve spectrum.
The emission lines, radiative recombination continuum, and
  bremsstrahlung emission produced by them
are modelled with the \texttt{pion} code\cite{Mehdipour16}.
In addition, we take into account
the Thomson-scattered component \cite[see e.g.,][]{McKaig23} of the direct emission
(\texttt{zcutoffpl}), which is not included in the \texttt{pion} model,
by setting its normalization
to be consistent with the column density and covering fraction of the
photoionized plasmas. Here we ignore self-absorption by bound electrons.
The detailed physical interpretation of this photoionized emission model will be discussed in a forthcoming paper (Guainazzi et al., in preparation).

We simultaneously fit the Resolve and NuSTAR spectra by adopting C-statistics 
for the former and $\chi^2$ statistics for the latter.
Our model successfully reproduced the data: for a total of 30 free
parameters, we obtain C-statistics $=2695.8$ for 2316 bins (Resolve),
$\chi^2 = 213.4$ for 232 bins (FPMA), and $\chi^2 = 216.4$ for 212
bins (FPMB). 
We confirm all AGN components introduced above are significantly required
  at $>99\%$ confidence limits.
Supplementary Fig.~4 plots the spectra of
Resolve and NuSTAR corrected for the effective area, the best-fit
models for the total and the AGN component, and the fitting residuals. A zoom of Fig.~\ref{fig:spec}, including only the energy range of the Fe K$\alpha$ and K$\beta$ fluorescent lines, is shown in Supplementary Fig.~5.
The best-fit parameters of the AGN component, together with the parameters of the Gaussian kernel
are summarized in Extended Data Table~1.
The error on each parameter corresponds to a 1$\sigma$ 
conﬁdence limit for a single parameter estimated by 
Markov Chain Monte Carlo method.
Supplementary Fig.~6 presents the corner plot among the metal abundances. 
The high inclination angle we obtain, $i=79.1\pm0.4$ degrees, is compatible with estimates from other wavelengths: $i>70$ degrees \cite{Tristram14} and $i>83$ degrees \cite{Isbell22} from mid-infrared interferometry, and $i\sim90$ degrees from H$_2$O maser observations showing evidence for a warped disk\cite{Greenhill03}.
The hydrogen column density of the 
foreground absorption in the host galaxy is found to be 
$N_{\rm H}^{\rm host} \simeq 2\times10^{22}$ cm$^{-2}$, which is consistent
with the previous estimate from the optical extinction \cite{Roche06}
assuming that the absorber has similarly high metallicity as in the torus
material and that dust mass is proportional to metal mass
(i.e., it has a 2.3 times larger dust-to-gas ratio than in the Galactic interstellar medium
where a hydrogen column density $N_{\rm H}$ is related to optical $V$-band extinction $A_V$ as $N_{\rm H} \approx 2\times10^{21}$ cm$^{-2}$ $A_V$).
We have confirmed that our results on the metal abundances
are little affected within the uncertainties in the fluxes of the
contaminating sources.
Since emission lines from the photoionized plasmas are spectroscopically separated from the neutral lines in the Resolve data, their effects on our elemental abundance measurements are ignorable.
  
We have determined the redshift parameters of the iron K$\alpha$ and
K$\beta$ lines (with respect to the \citet{Hoelzer97} values) to be
$z=0.00155 \pm 0.00001 (\pm 0.00005)$ and $z=0.00158\pm 0.00004 (\pm
0.00005)$, respectively, after applying a heliocentric correction of $20$ km s$^{-1}$
(i.e., for the orbital motion of the Earth around the Sun; 
here we have ignored the motion of the satellite relative to the Earth, which has a velocity amplitude of 8 km s$^{-1}$ with a period of 90 minutes and is averaged out during the observation.)
The errors in the
parenthesis denote a gain calibration uncertainty of 0.3 eV (Resolve team, private communication).
Thus, our results are consistent with the host galaxy redshift,
$z=0.001448 \pm 0.000010$, measured by H~I observations \cite{Koribalski04}.
For reference,
Supplementary Table~1
summarizes the line centroid energies of major fluorescence
lines in the literature based on ground experiments or theoretical
calculations \cite{Hoelzer97,Ito16,Ito18,Bearden67}. Note that the line energy depends on the chemical state of atoms
by an order of 1 eV for iron K$\alpha$ \cite[e.g.,][]{Baydas12}, 
which produces an additional systematic uncertainty in discussing the Doppler shift of the line emitter.
  Thus, a conservative limit on the line-of-sight velocity of the torus relative to the galaxy based on iron K$\alpha$ is $-16$ to $+78$ km s$^{-1}$ (the plus sign corresponds to a receding motion from us).

\section{Possible Contamination in Neutral Line Flux from Diffuse Emission}\label{sec:diffuse}

In our best-fit model, the spectrum of the CGX1+diffuse component is
modelled by a power law with a high energy cutoff, and hence does not
include any emission lines. However, if the diffuse emission around
the nucleus contained a scattered component of the AGN emission off
cold gas, then it could contain fluorescence lines contaminating the
line fluxes from the nucleus in the Resolve spectrum.

To evaluate its possible impact on the measurement of metal
abundances of the torus, we estimate maximum fluxes of the fluorescence
lines from the diffuse emission, 
using the continuum flux of the CGX1+diffuse emission
and theoretically expected equivalent widths in the scattered component from low column-density gas.
Since the relative contribution from the diffuse
emission becomes important at soft energies, here we focus on Ar
K$\alpha$ and Ca K$\alpha$ lines.
We have confirmed that the flux of
diffuse emission in the 2.5--4.0 keV detected with Chandra
is consistent with that inferred from our best-fit spectral model.
Utilizing the Monte-Carlo based radiative transfer code SKIRT \cite{VanderMeulen23}, we calculate the X-ray reflection spectrum from neutral,
constant density gas in a spherical geometry irradiated by the central
source. We assume a power law spectrum with a photon index of 1.9 as
the intrinsic emission, the chemical abundances in Extended Data
Table~1, and a column density of $10^{20}$ cm$^{-2}$
integrated from the centre to the outer radius of the sphere, which is
necessary to account for the diffuse flux by Thomson scattering. This
calculation yields the expected equivalent widths of 31 eV for Ar K$\alpha$ and
36 eV for Ca K$\alpha$.
By multiplying the continuum flux of the CGX1+diffuse emission, we find
that these 
correspond to fractions of 16\% (Ar K$\alpha$) and 10\% (Ca K$\alpha$)
in the total line fluxes observed with Resolve,
which are comparable to the 1$\sigma$ statistical uncertainties.
These numbers should be taken as upper limits, because we assume that the CGX1+diffuse components is fully attributed to a scattered component from cold gas, which is likely not to be the case; 
in fact, 
the Chandra spectrum of the diffuse emission within a radius of 6 arcsec around the nucleus shows no significant Ar K$\alpha$ line with a 90\% upper limit on the equivalent width of 14 eV, a factor of 0.45 smaller than the above calculation.
Thus, we conclude that our results on the metal
abundances are robust. Note that the possible contamination from the
diffuse emission, if any, works to strengthen our arguments on the
sub-solar abundance ratios of the Ar/Fe and Ca/Fe.

\section{Effects by Adopting Different Torus Geometry}\label{sec:torusgeometry}

To ensure that our results on the metal abundances do not depend on
detailed spectral modelling of the torus reflection component, here we
examine the effects by adopting different torus geometry. Here we focus
on (1) the size and number of clumps in the XCLUMPY model and (2)
different spatial distribution of clumps from that in XCLUMPY.

\subsection{Size and Number of Clumps in XCLUMPY}

To examine the effect of adopting a different clump size from $R_{\rm clump}=$0.002 pc
(default value), we calculate
the reflection spectra by adopting $R_{\rm clump}=$0.001 pc or 0.005
pc by keeping the spectral parameters at the best-fit values in
Extended Data Table~1. The results are plotted in Supplementary Fig.~7. As noticed, the spectra with different clump sizes
in the 6.0--6.6 keV band are almost identical one another, including
the line intensity and Compton shoulder fraction, which are important
to constrain the iron to hydrogen abundance ratio (see Main
Text). Next, we change the number of clumps along the sight line on
the equatorial plane, $N_{\rm clump}^{\mathrm Equ}$, from 10 (default) to 5 or
15. We confirm that the resultant spectra in the 6.0--6.6 keV band
again show no notable differences among
them. Thus, we conclude that our results on the chemical abundances
are robust against choices of $R_{\rm clump}$ and $N_{\rm
  clump}^{\mathrm Equ}$ in the XCLUMPY geometry.

\subsection{Clump Distribution}\label{sec:ixclumpy}

To examine if the torus reflection spectra sensitively depend on the
clump configuration, here we consider a different torus geometry from
that in XCLUMPY characterized by a circular-sector cross-section
(Supplementary Fig.~3(b)). The
geometry is similar to those adopted in the smooth torus models
eTORUS\cite{Ikeda09} and BORUS\cite{Balokovic18}.  The torus structure
is defined by the following parameters: the torus angular width
$\theta$, the number of clumps along the equatorial plane
$N_\mathrm{clump}^\mathrm{Equ}$, the equatorial column density
$N_\mathrm{H}^\mathrm{Equ}$, and the inner and outer radii of the
torus, $r_\mathrm{in}$ and $r_\mathrm{out}$, respectively.  Following
the configuration of the XCLUMPY model, we adopted $r_\mathrm{in} =
0.05$, $r_\mathrm{out} = 1$, and $N_\mathrm{clump}^\mathrm{Equ} = 10$
for our simulations. The clumps are distributed uniformly and
randomly.

The number density function, $d(r, \phi, z)$ [pc$^{-3}$], is defined in a cylindrical coordinate system (where $r$ is the radius, $\phi$ is the azimuth, and $z$ is the height) as a constant value $N$:
\begin{equation}
    d(r, \phi, z) = N,
\end{equation}
where $N$ is the normalization factor. This normalization is determined from the number of clumps along the equatorial plane, $N_\mathrm{clump}^\mathrm{Equ}$, via the relation:
\begin{align}
    N_\mathrm{clump}^\mathrm{Equ} &= \int_{r_\mathrm{in}}^{r_\mathrm{out}} d(r, 0, 0) \, \pi R_\mathrm{clump}^2 \, dr, \\
    N &= \frac{N_\mathrm{clump}^\mathrm{Equ}}{\pi R_\mathrm{clump}^2 (r_\mathrm{out} - r_\mathrm{in})}.
\end{align}
The total number of clumps in the torus, $N_\mathrm{clump}^\mathrm{Tot}$, is obtained by integrating the number density function over the torus volume:
\begin{align}
    N_\mathrm{clump}^\mathrm{Tot} &= \int_{- (r_\mathrm{out} - r_\mathrm{in}) \sin{\theta}}^{(r_\mathrm{out} - r_\mathrm{in}) \sin{\theta}} \int_{0}^{2\pi} \int_{r_\mathrm{in} + |z| \cot{\theta}}^{r_\mathrm{in}+\sqrt{(r_\mathrm{out} - r_\mathrm{in})^2 - z^2}} d(r, \phi, z) \, r \, dr \, d\phi \, dz \nonumber \\
    &= \frac{N_\mathrm{clump}^\mathrm{Equ} (r_\mathrm{out} - r_\mathrm{in}) \left( 4(r_\mathrm{out} - r_\mathrm{in})\sin{\theta} + 6 r_\mathrm{in} \theta \right)}{3 R_\mathrm{clump}^2}.
\end{align}

We obtained a C-statistic of 2816.4 (Resolve; 2316 bins) and $\chi^2$
of 208.4 (FPMA; 232 bins), and 210.6 (FPMB; 212 bins), with 24 free
parameters.
This model yields the following parameters: a photon index of $\Gamma
= 1.92\pm0.05$, $\theta = 13.5\pm0.7$ degrees, $i = 78.2\pm0.7$
degrees, and an iron abundance of $Z(\mathrm{Fe}) =
2.21^{+0.03}_{-0.04}$ solar.  Notably, the derived iron abundance is
consistent with that obtained using the XCLUMPY model ($Z(\mathrm{Fe})
= 2.26\pm0.04$). This indicates that our results on the chemical
abundances are not largely affected by the torus geometry adopted.

\section{Metal Enrichment Models}\label{sec:snmodels}

As described in the main text, our benchmark model to interpret the
observed abundance pattern consists of about 90\% CCSNe with progenitors
less massive than $20 M_\odot$ and about 10\% SNe~Ia.
We adopted the SN model by \citet{Nomoto13} for CCSNe and those by
\citet{Fink14} and \citet{Shen18} for near-and sub-$M_\textup{Ch}$
white dwarfs, respectively, assuming one-solar initial metallicity.
We used the Salpeter IMF \cite{Salpeter55} to integrate the
initial-mass dependent CCSN yields, ignoring the contribution from
hypernova. To ensure that our main results (i.e., the high CCSN
fraction and the presence of the upper mass limit for the progenitors
of CCSNe) are robust against these assumptions, here we examine the
effects by adopting other model settings.

\subsection{SN~Ia Models}

For the SN~Ia model, we test two shock-propagation models in an
exploding white dwarf: pure-deflagration and delayed-detonation. The
3D hydrodynamical simulation utilized in \citet{Fink14} assumes the
pure-deflagration scenario, whereas that in \citet{Seitenzahl13b}
adopts the delayed-detonation one. These two works also examine three
initial core densities for yield calculation: normal, high, and
low. Accordingly, we try all six settings to fit our metal abundance
pattern. For each setting, we adopt five different mass cutoffs for
CCSNe progenitors. Supplementary Fig.~8 plots
$\chi^2$/d.o.f. values against the estimated SN~Ia fractions. The
upwards and downwards triangles represent the calculation with
\citet{Seitenzahl13b} and with \citet{Fink14}, respectively.

Most of them show a similar trend that the small upper mass limit
provides a small fraction of SNe~Ia and a better $\chi^2$.
Lower SN~Ia fractions than the representative case of Perseus
($15$--$38$\%, \cite{Simionescu19}, shown by grey shaded area in Supplementary Fig.~8)
can only be achieved with the most severe upper limit of $20 M_\odot$.
As for the densities of the ignition core, the normal density (approximately
$3 \times 10^9$\,g\,cm$^{-3}$)
is preferred for pure deflagration, and the normal and high densities can be acceptable for delayed detonation.
Despite the comprehensive analysis of galaxy clusters, it remains unclear
which of pure deflagration or delayed detonation predominantly contributes to SN~Ia enrichments in the universe.
Hence, we do not conclude which SN~Ia scenario is better based on
our fitting results; instead,
the difference between yields with pure-deflagration and delayed-detonation for a normal density dwarf
is taken into account as uncertainty $\sigma_\textup{Ia}$ in the best-fitting model (Fig.~\ref{fig:cut}).

\subsection{CCSN Models}

\subsubsection{Alternative Yield Calculation}

Alternative CCSN calculations by \citet{Sukhbold16} are examined with
both the pure-deflagration and delayed-detonation SNe~Ia (pluses and
crosses in Supplementary Fig.~8).
We find that the classical calculations by \citet{Nomoto13} generally result in
better fits and more reasonable fractions than \citet{Sukhbold16}.
This trend is similar to that found by \citet{Mao21} for the starburst
galaxy Arp~299. Thus, we adopt the yield of \citet{Nomoto13} as our
benchmark for enrichment modelling.

\subsubsection{IMF}

We adopt the IMF-weighted yields of CCSNe using the top-heavy
index of $-1.0$, considering the possibility that 
the IMF in starburst systems may follow a top-heavy behaviour
rather than the Salpeter index of $-2.3$ \cite[e.g.,][]{Bastian10}.
The results are given in Supplementary Fig.~8 with
vertical rectangles. As noticed, the fraction of SNe~Ia increases
compared with the Salpeter case; 
this is naturally interpreted
because massive progenitors make a greater contribution
to metal enrichment with the top-heavy IMF than with the baseline case.
By contrast, 
the bottom-heavy index of $-3.0$ slightly reduces the SN~Ia fractions
as shown by horizontal rectangles in Supplementary Fig.~8,
indicating that reducing the contribution from massive progenitors is essential to reproduce the observed abundance pattern. 
However, the bottom-heavy IMF model suppresses the SN~Ia contribution less effectively
than the assumption of the CCSN model with an upper mass limit.

\subsubsection{Hypernova Contribution}

We consider a possible contribution from the hypernova (HN) channel in
massive stars \cite{Nomoto13}. To take HNe into account, we replace
the CCSN yield tables for the massive progenitor regime ($> 20
M_\odot$) with the HN yields from \citet{Nomoto13}. The wherewithal to
calculate the IMF-weighted yield follows the same strategy as
described in the main text.  The circles in Supplementary
Fig.~8 represent the results of this CCSN+HN
scenario.  While all cases show slightly lower SN~Ia fractions with
better $\chi^2$ values than the pure CCSN case, the estimated
values are still comparable to the Perseus environment; thus, this
model would not be an alternative to the mass cutoff modelling.

\subsubsection{Progenitor Metallicity}

Our measurement of the Fe/H ratio reveals super-solar metallicity
(about 2\,solar) in the Circinus centre. If this result is directly
interpreted as metal-rich environments in starbursting regions, it
motivates us to test high initial metallicity for CCSN
progenitors. Since \citet{Nomoto13} extends their yield calculations
to the super-solar metallicity regime, we adopt the CCSN yield tables
of $Z_\textup{CCSN} = 0.05$ to replicate the benchmark modelling with
white dwarf progenitors \cite{Fink14, Shen18}. The
resulting enrichment properties are summarized in
Table~\ref{tab:frac} for the progenitor mass upper limits of 40
$M_\odot$, 25 $M_\odot$, and 20 $M_\odot$. As noticed, the super-solar
metallicity progenitors also prefer the mass cutoff CCSN model with a
lower SN~Ia fraction less than 10\%, which is consistent with the
benchmark scenario. Interestingly, the $\chi^2$ values with
$Z_\textup{CCSN} = 0.05$ are slightly better than with
$Z_\textup{CCSN} = 0.02$, implying that the bulk of CCSNe in the
Circinus centre are from progenitors formed in a metal-rich
environment. However, we keep our benchmark model at solar metallicity
for a fair comparison, and include the difference between the
abundances with $Z_\textup{CCSN} = 0.02$ and $Z_\textup{CCSN} = 0.05$
as uncertainty $\sigma_\textup{CC}$ in the best-fitting model
(Fig.~\ref{fig:cut}).

\section{Mass Transfer in Circinus Centre}\label{sec:masstransfer}

Here we show that the gas in the torus region 
must be transferred from outer regions on the basis of our
metallicity measurement.
Assuming the mass
accretion rate of $0.3 M_\odot$ yr$^{-1}$ at $r=0.27$
pc \cite{Izumi23}, the star formation rate of
$0.16$ $M_\odot$ yr$^{-1}$
within 15 pc$\times$15 pc \cite{Davies07}, and
the total gas mass of about $10^6 M_\odot$ \cite{Wada16} within
15 pc$\times$15 pc,
the depletion
timescale (or replacement timescale assuming a steady state) is
estimated to be $2\times10^6$ yr. During this period a total of $3\times10^3$
SNe take place in the torus and produce an iron mass of 200
$M_\odot$, corresponding to about 0.1 solar relative to the hydrogen
mass. This is not sufficient to explain the observed metallicity in
the Circinus centre
(Fe/H = 2.3 solar), and therefore metal-enriched gas must be transferred from
the outer disk, for example, from $r\sim200$ pc 
where a young starburst region was identified \cite{Marconi94}.

\clearpage

\noindent
\textbf{Data availability}
The observational data analysed during this study are available at
NASA's High Energy Astrophysics Science Archive Research Center
(HEASARC; https://heasarc.gsfc.nasa.gov/) 
with observation IDs 000162000 (XRISM), 60901013002 (NuSTAR), 
0932990101 (XMM-Newton), and 10223, 10224, 10225, 10226, 10832,
10833, 10842, 10843, 10844, 10850, 10872, and 10873 (Chandra).
Line energies
in the NIST atomic spectra database are available online (https://www.nist.gov/pml/atomic-spectra-database).

\bigskip
\noindent
\textbf{Code availability}
The codes used for the data reduction are available from the HEASARC
website (https://heasarc.gsfc.nasa.gov/docs/software/heasoft) and the
ESA’s website (https://www.cosmos.esa.int/web/xmm-newton/sas). The
spectral fitting tools are freely available online (https://heasarc.gsfc.nasa.gov/xanadu/xspec for XSPEC and https://zenodo.org/records/17313851 for SPEX).

\bigskip
\noindent
\textbf{Acknowledgements}
We thank Aiko Miyamoto for her help in analysing the Chandra imaging data.
This work was supported by JSPS KAKENHI grant numbers JP22H00158, JP22H01268, JP22K03624, JP23H04899, JP21K13963, JP24K00638, JP24K17105, JP21K13958, JP21H01095, JP23K20850, JP24H00253, JP21K03615, JP24K00677, JP20K14491, JP23H00151, JP19K21884, JP20H01947, JP20KK0071, JP23K20239, JP24K00672, JP24K17104, JP24K17093, JP20K04009, JP21H04493, JP20H01946, JP23K13154, JP19K14762, JP20H05857, JP23K03459, JP24KK0070, JP24H01810, JP23K13153, JP24K00673, JP25H00672, and JP22KJ1990, and NASA grant numbers 80NSSC23K0650, 80NSSC20K0733, 80NSSC18K0978, 80NSSC20K0883, 80NSSC20K0737, 80NSSC24K0678, 80NSSC18K1684, and 80NNSC22K1922. Y.~Ueda acknowledges the support from the Kyoto University Foundation.
E.~B. was supported by the Israel Science Foundation (grant No. 2617/25). L.~C. acknowledges support from NSF award 2205918. C.~D. acknowledges support from STFC through grant ST/T000244/1. L.~Gu acknowledges financial support from Canadian Space Agency grant 18XARMSTMA. A.~Tanimoto and the present research are in part supported by the Kagoshima University postdoctoral research program (KU-DREAM). Satoshi Yamada acknowledges support by the RIKEN SPDR Program. I.~Z. acknowledges partial support from the Alfred P. Sloan Foundation through the Sloan Research Fellowship. M.~Sawada acknowledges the support by the RIKEN Pioneering Project Evolution of Matter in the Universe (r-EMU) and Rikkyo University Special Fund for Research (Rikkyo SFR). N.~W. and T.~P. acknowledge the financial support of the GAČR EXPRO grant No. 21-13491X. Part of this work was performed under the auspices of the U.S. Department of Energy by Lawrence Livermore National Laboratory under Contract DE-AC52-07NA27344. The material is based upon work supported by NASA under award number 80GSFC21M0002. This work was supported by the JSPS Core-to-Core Program, JPJSCCA20220002. The material is based on work supported by the Strategic Research Center of Saitama University. This work made use of the JAXA Supercomuter System Generation 3 (JSS3).

\bigskip
\noindent
\textbf{Author contributions}
As the leader of the Circinus galaxy target team in the XRISM Science
Team (XST), Y.~Ueda led this research project and wrote the
manuscript. R.U. and S.O. led the data analysis, and
K.~Fukushima led the interpretation of the results, all contributing
to prepare the manuscript. M.G. and T.~Yaqoob analysed the
XRISM data, and A.~Tanimoto reduced the NuSTAR data. K.~Maeda,
Y.~Fujita, R.M., K.~Matsushita, F.P., T.I.,
K.~Fujiwara, and Y.N. discussed the results. F.S.P., C.K., and
M.E.E. contributed to the data calibration of Resolve and wrote
the related section in Methods. L. Gallo, E.K., Satoshi Yamada,
J.M.M., T.~Kawamuro., Y.~Terashima, B.V.M., Y.~Fukazawa,
E.B., and L.C. provided comments on the content of the
manuscript, and helped to improve the manuscript. The science goals of
XRISM were discussed and developed over 7 years by the XST, all
members of which are authors of this manuscript. All the instruments
were prepared by the joint efforts of the team. The manuscript was
subject to an internal collaboration-wide review process. All authors
reviewed and approved the final version of the manuscript.

\bigskip
\noindent
\textbf{Competing interests}
The authors declare no competing interests.

\clearpage

\begin{table}[ht]
\centering
\caption{{\bf The best-fitting number fraction of each type of SN to reproduce the observed abundance ratios of Ar/Fe, Ca/Fe, Cr/Fe, Mn/Fe, and Ni/Fe in the Circinus centre.} Different assumptions on the upper mass limits for CCSN progenitors are adopted.}\label{tab:frac}
\begin{tabular}{ccccc} \hline\hline
  $M_\textup{CCSN, up}$$^a$ & \multicolumn{2}{c}{SN~Ia (\%)} & CCSN (\%) & Reduced $\chi^2$\\
 & near-$M_\textup{Ch}$$^b$ & sub-$M_\textup{Ch}$$^c$ & &  (Degrees of freedom)\\ \hline
 \multicolumn{5}{c}{$Z_\textup{CCSN}=0.02$ (solar metallicity)} \\
 40$M_\odot$ & 19 & 21 & 60 & 3.32 (3)\\
 35$M_\odot$ & 18 & 19 & 63 & 3.14 (3)\\
 30$M_\odot$ & 17 & 17 & 66 & 2.73 (3)\\
 25$M_\odot$ & 13 & 9 & 78 & 1.90 (3)\\
 20$M_\odot$ & 8 & 0 & 92 & 1.12 (3)\\
 18$M_\odot$ & 9 & 1 & 90 & 1.60 (3)\\\hline
  \multicolumn{5}{c}{$Z_\textup{CCSN}=0.05$ (super-solar metallicity)} \\
 40$M_\odot$ & 14 & 22 & 64 & 2.90 (3)\\
 25$M_\odot$ & 8 & 9 & 73 & 1.47 (3)\\
 20$M_\odot$ & 5 & 3 & 92 & 1.06 (3)\\
 18$M_\odot$ & 5 & 2 & 93 & 1.16 (3)\\ \hline
\end{tabular}
\begin{flushleft}
  $^{a}$ Upper mass limit for CCSN progenitors.\\
  $^{b}$ SN Ia from near-$M_\textup{Ch}$ progenitors.\\
  $^{c}$ SN Ia from sub-$M_\textup{Ch}$ progenitors.\\
\end{flushleft}
\end{table}

\clearpage

\noindent
\textbf{XRISM Collaboration}

\vspace{3mm}

\noindent
\noindent
XRISM Collaboration$^{1}$,  
Marc Audard$^{2}$,  
Hisamitsu Awaki$^{3}$,  
Ralf Ballhausen$^{4,5,6}$,  
Aya Bamba$^{7}$,  
Ehud Behar$^{8}$,  
Rozenn Boissay-Malaquin$^{9,5,6}$,  
Laura Brenneman$^{10}$,  
Gregory V.\ Brown$^{11}$,  
Lia Corrales$^{12}$
% ...

\vspace{3mm}
\noindent
$^1$Corresponding Authors: Y. Ueda (ueda@kusastro.kyoto-u.ac.jp),
R. Uematsu (uematsu@kusastro.kyoto-u.ac.jp), S. Ogawa (sogawa@ac.jaxa.jp), and K. Fukushima (fukushima.kotaro@jaxa.jp).
$^2$Department of Astronomy, University of Geneva, Versoix CH-1290, Switzerland,  
$^3$Department of Physics, Ehime University, Ehime 790-8577, Japan,  
$^4$Department of Astronomy, University of Maryland, College Park, MD 20742, USA,  
$^5$NASA / Goddard Space Flight Center, Greenbelt, MD 20771, USA,  
$^6$Center for Research and Exploration in Space Science and Technology, NASA / GSFC (CRESST II), Greenbelt, MD 20771, USA,  
$^7$Department of Physics, University of Tokyo, Tokyo 113-0033, Japan,  
$^8$Department of Physics, Technion, Technion City, Haifa 3200003, Israel,  
$^9$Center for Space Science and Technology, University of Maryland, Baltimore County (UMBC), Baltimore, MD 21250, USA,  
$^{10}$Center for Astrophysics | Harvard-Smithsonian, MA 02138, USA,  
$^{11}$Lawrence Livermore National Laboratory, CA 94550, USA,  
$^{12}$Department of Astronomy, University of Michigan, MI 48109, USA,  
% ...

\clearpage

\section*{Extended Data}
\label{fig:ext} 

\renewcommand{\figurename}{Extended Data Fig.}
\setcounter{figure}{0}
\renewcommand{\thefigure}{\arabic{figure}}
\renewcommand{\theHfigure}{ED\arabic{figure}}

\renewcommand{\tablename}{Extended Data Table}
\setcounter{table}{0}
\renewcommand{\thetable}{\arabic{table}}
\renewcommand{\theHtable}{ED\arabic{table}}
 
\begin{figure}[ht]
\centering
\includegraphics[width=\linewidth]{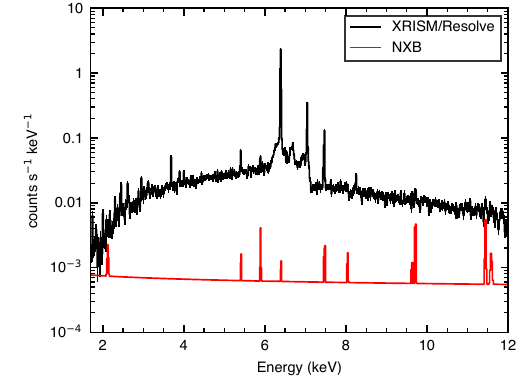}
\caption{{\bf
    Contribution of the non X-ray background (NXB) in the Circinus spectrum.}
    The red line denotes the modelled NXB spectrum. The black data points are
    the observed spectrum of Circinus (including the NXB) with error
    bars denoting the 1$\sigma$ confidence limits in photon
    statistics.
}\label{fig:nxb}
\end{figure}

\begin{figure}[ht]
\centering
\includegraphics[width=\linewidth]{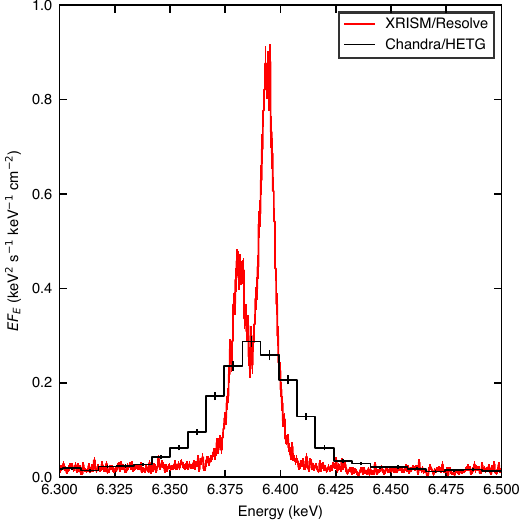}
\caption{
{\bf 
  Comparison of the XRISM/Resolve spectrum and Chandra/HETG first order spectrum around the iron K$\alpha$ line of Circinus.} The former (this work) is binned with 0.5 eV. The latter is adopted from \citet{Uematsu21} and is binned with 9 eV.
The error bars denote the 1$\sigma$ confidence limits in photon statistics.
}
\label{fig:heg}
\end{figure}

\begin{figure}
\centering
\includegraphics[width=\columnwidth]{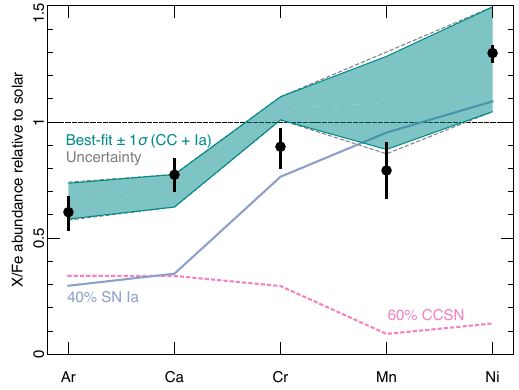}
\caption{
{\bf 
 Observed elemental abundances of the Circinus centre and theoretically calculated values with an upper mass limit of 40 $M_\odot$for CCSN progenitors.}
Relative abundances with respect to iron determined with XRISM (X/Fe, filled circles with error bars denoting 1$\sigma$ confidence limits estimated by
Markov Chain Monte Carlo method) normalized to the corresponding proto-solar values (horizontal dashed line).
The green shaded area represents the 1$\sigma$ uncertainty region of the best-fit model
  consisting of \textcolor{magenta}{($60\pm10$)\%} CCSNe from progenitor stars with $<40 M_\odot$, \textcolor{magenta}{($19\pm7$)\%} near-$M_\textup{Ch}$ SNe~Ia, and \textcolor{magenta}{($21\pm11$)\%} 
sub-$M_\textup{Ch}$ SNe~Ia in number fraction, whose 
contributions to X/Fe are represented by the magenta, blue, and purple curves, respectively.
  Thin grey-dashed lines show the total uncertainty, including differences in the predicted yields by alternative SN~Ia and CCSN modelling as 
 $\sqrt{\sigma_\textup{stat}^2 + \sigma_\textup{Ia}^2 + \sigma_\textup{CC}^2}$ (see Methods Section~\ref{sec:snmodels}).
}
\label{fig:nocut}
\end{figure}

\clearpage
\renewcommand{\arraystretch}{1.3}

\begin{table}
  \begin{center}
    \caption{{\bf Best-fit Parameters of XCLUMPY.} Errors denote the 1$\sigma$ confidence limits.}
    \label{tab:agnparameter}
    \begin{tabular}{ccc}
      \hline\hline
      Parameter&Value&Units \\ \hline
      $Z$$^{a}$&$1.71\pm0.04$ &solar\\
      $Z$(Ar)$^{b}$ &$1.39_{-0.20}^{+0.16}$&solar\\
      $Z$(Ca)$^{b}$ &$1.75\pm0.17$&solar\\
      $Z$(Cr)$^{b}$ &$2.02_{-0.22}^{+0.18}$&solar\\
      $Z$(Mn)$^{b}$ &$1.79\pm0.28$&solar\\
      $Z$(Fe)$^{b}$ &$2.26\pm0.04$&solar\\
      $Z$(Ni)$^{b}$ &$2.94_{-0.11}^{+0.09}$&solar\\
      $N_\mathrm{H}^\mathrm{equ}$ &$2.26_{-0.16}^{+0.17}\times10^{25}$ & cm$^{-2}$ \\
      $\sigma$$^{c}$& $9.2\pm0.3$ & degree \\
      $i$&$79.1\pm0.4$&degree \\
      $\Gamma$$^{d}$&$1.90_{-0.04}^{+0.05}$& \\
      $E_{\rm cut}$$^{e}$&$65_{-5}^{+7}$&keV \\
      $A$$^{f}$&$1.51_{-0.16}^{+0.18}$& photons cm$^{-1}$ s$^{-1}$ keV$^{-1}$\\
      $L_{2-10}$$^{g}$&$8.84_{-0.54}^{+0.63}\times10^{42}$& erg s$^{-1}$\\
      $N_{\rm H}^{\rm host}$$^{h}$&$1.66_{-0.15}^{+0.16}\times10^{22}$ & cm$^{-2}$\\
      line width$^{i}$ &$210\pm10$& km s$^{-1}$\\\hline
   \end{tabular}
\begin{flushleft}
  $^{a}$ Abundance of other metals than Ar, Ca, Cr, Mn, Fe, and Ni, relative to hydrogen.\\
  $^{b}$ Abundance of the element in parenthesis relative to hydrogen.\\
  $^{c}$ Torus angular width.\\
  $^{d}$ Photon index of the intrinsic AGN component.\\
  $^{e}$ Cutoff energy of the intrinsic AGN component.\\
  $^{f}$ Normalization at 1 keV of the intrinsic AGN component.\\
  $^{g}$ Intrinsic AGN luminosity in the 2--10 keV band (a distance of 4.2 Mpc is assumed).\\
  $^{h}$ Hydrogen column density of the foreground absorption in the host galaxy.\\
  $^{i}$ FWHM of the Gaussian kernel in velocity.\\
\end{flushleft}
  \end{center}
\end{table}

\clearpage

\section*{Supplementary Information}

\renewcommand{\figurename}{Supplementary Fig.}
\setcounter{figure}{0}
\renewcommand{\thefigure}{\arabic{figure}}
\renewcommand{\theHfigure}{S\arabic{figure}}

\renewcommand{\tablename}{Supplementary Table}
\setcounter{table}{0}
\renewcommand{\thetable}{\arabic{table}}
\renewcommand{\theHtable}{S\arabic{table}}

\begin{figure}[ht]
\centering
\includegraphics[width=1.0\linewidth]{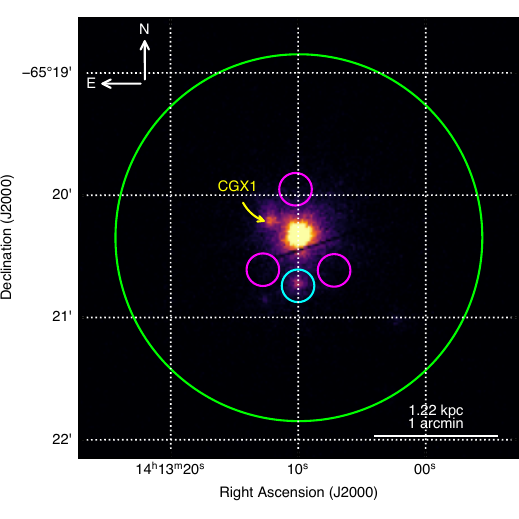}
\caption{{\bf   
The XMM-Newton/MOS1 image in the 0.2--12 keV band around the Circinus centre.} The small cyan circle and three magenta circles are the source and background extraction regions for the ``XMM-CGX2'' spectrum. The large green circle corresponds to the source region for the ``XMM-all'' (pn) spectrum. 
}\label{fig:mos}
\end{figure}

\newpage
\begin{figure}[ht]
\centering
\includegraphics[width=1.0\linewidth]{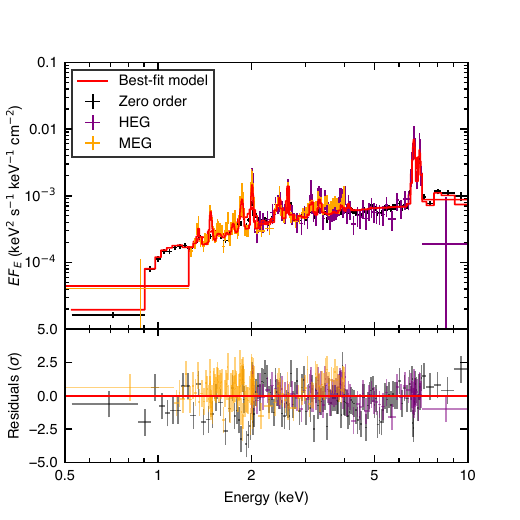}
\caption{{\bf The Chandra/HETG spectra of CGX2 folded with the energy response overlaid with the best-fit model.} The black, purple, and orange colours correspond to
  the spectra of zero order, HEG, and MEG, respectively. The error bars denote the 1$\sigma$ confidence limits in photon statistics.
The red line represents the best-fit model.
The lower panel shows the fitting residuals normalized by the 1$\sigma$ errors.
}\label{fig:cgx2chan}
\end{figure}

\newpage
\begin{figure}[ht]
\centering
\includegraphics[width=1.0\linewidth]{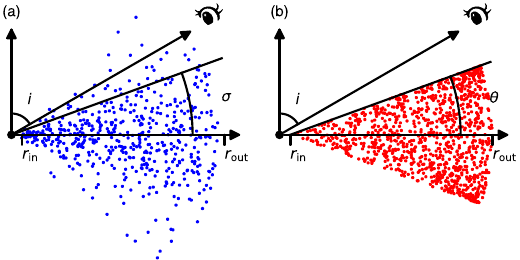}
\caption{{\bf Cross section views of the distribution of
    clumps in our torus models.} (a) That in the XCLUMPY model. (b) That in the torus model described in Methods Section~\ref{sec:ixclumpy}.
}\label{fig:xclumpy}
\end{figure}

\newpage
\begin{figure}[ht]
\centering
\includegraphics[width=1.0\linewidth]{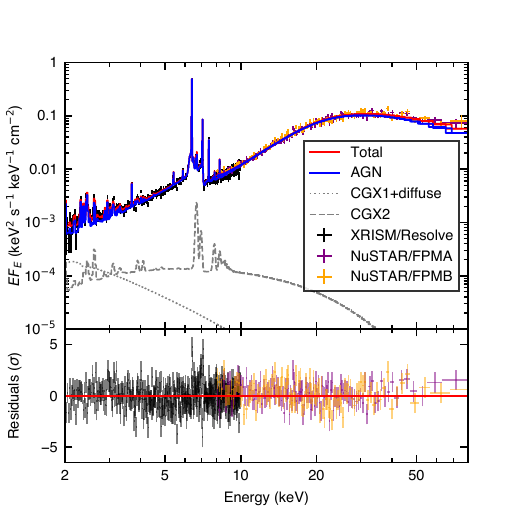}
\caption{{\bf The XRISM/Resolve and NuSTAR spectra of Circinus.} They are folded by the energy responses in units of energy flux (i.e., corrected for the effective area). The black, purple, and orange colours correspond to
the spectra of XRISM/Resolve, NuSTAR/FPMA, and NuSTAR/FPMB, respectively.
The error bars denote the 1$\sigma$ confidence limits in photon statistics. 
The red curve represent the best-fit model. 
The blue solid, grey dashed, grey dotted curves represent the AGN, CGX2, and CGX1+diffuse components,
respectively.
The lower panel shows the fitting residuals normalized by the 1$\sigma$ errors.
  }\label{fig:broad}
\end{figure}

\newpage
\begin{figure}[ht]
\centering
\includegraphics[width=1.0\linewidth]{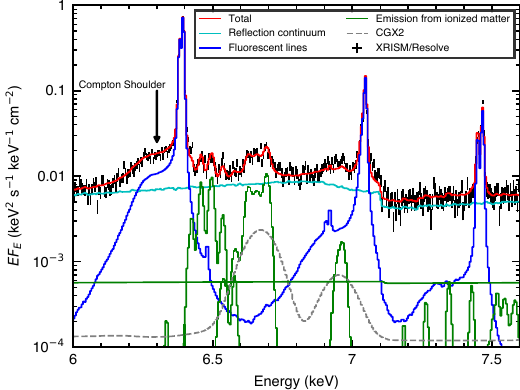}
\caption{
  {\bf The XRISM/Resolve spectrum of Circinus in the 6---7.6 keV band overlaid with the best-fit model.} They are folded by the instrumental response in units of energy flux. The observed data are plotted in black (with error bars denoting the 1$\sigma$ confidence limits in photon statistics), and the total model is shown in red.
  The cyan and blue lines represent the reflection continuum and fluorescence lines, respectively,  
modelled by XCLUMPY. The green ones denote emission from ionized matter
modelled by \texttt{pion} (Guainazzi et al. in prep.) 
and the Thomson-scattered component. 
The grey dashed line shows the contribution from CGX2.
}\label{fig:feband}
\end{figure}

\newpage
\begin{figure}[ht]
\centering
\includegraphics[width=1.0\linewidth]{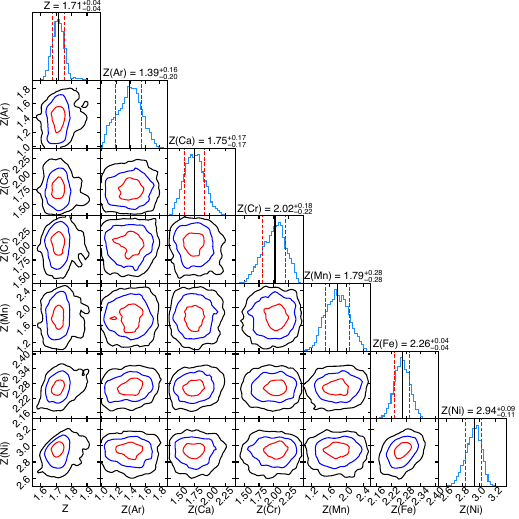}
\caption{
{\bf 
Posterior distributions of the elemental abundance ($Z$, $Z$(Ar),
$Z$(Ca), $Z$(Cr), $Z$(Mn), $Z$(Fe), $Z$(Ni)) obtained from the Markov Chain Monte Carlo analysis.} The contours represent the $1\sigma$, $2\sigma$, and
$3\sigma$ confidence levels from the inside out. The diagonal panels
show the one-dimensional marginalized distributions with the median
(black) and $1\sigma$ confidence intervals (red).
}\label{fig:corner}
\end{figure}

\newpage
\begin{figure}[ht]
\centering
\includegraphics[width=1.0\linewidth]{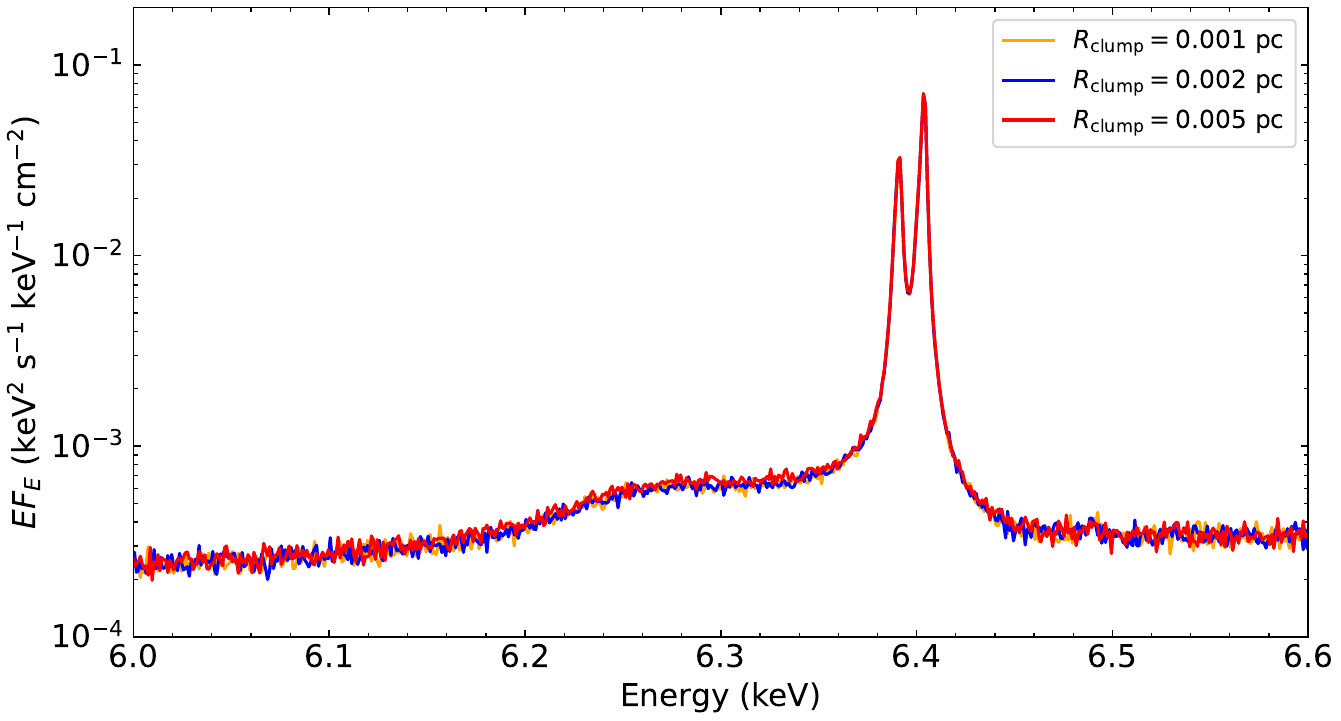}
\caption{
{\bf Comparison of simulated spectra in the 6.0--6.6 keV band calculated with different clump sizes in XCLUMPY.} We adopt $R_{\rm clump}=$0.001 pc, 0.002 pc (default), and 0.005 pc. The other parameters are the same as those in Extended Data Table~\ref{tab:agnparameter}.
}\label{fig:rclump}
\end{figure}

\newpage
\begin{figure}
\centering
\includegraphics[width=1.0\columnwidth]{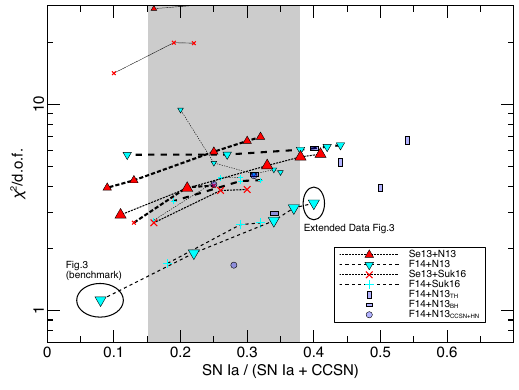}
\caption{
{\bf The derived $\chi^2$/d.o.f. values from the fit to the observed
abundance pattern plotted against the estimated SN~Ia fraction.}
The upward triangles represent the scenario consisting of CCSNe
\cite{Nomoto13} and delayed-detonation near-$M_\textup{Ch}$ SNe~Ia
\cite{Seitenzahl13b}. The line widths show the core density of
exploding white dwarfs (normal; $3 \times 10^9$, thin; $1 \times
10^9$, thick; $5 \times 10^9$\,g\,cm$^{-3}$).
The results with mass upper limits of $20$, $25$, $30$,
$35$, and $40 M_\odot$ (from left to right) for CCSN progenitors are
plotted. The downward triangles indicate the results with CCSNe
\cite{Nomoto13} and pure-deflagration near-$M_\textup{Ch}$ SNe~Ia
\cite{Fink14}, where the normal core density case of SNe~Ia
constitutes our benchmark results presented in the main text,
Fig.~\ref{fig:cut} and Extended Data Fig.~\ref{fig:nocut}.
The crosses and pluses show the results of the CCSN yields of \citet{Sukhbold16}
with white dwarf progenitors of \citet{Seitenzahl13b} and \citet{Fink14}, respectively.
The upper mass limits of CCSN progenitors $20$, $25$, and $40 M_\odot$
are applied from left to right.
The horizontal and vertical rectangles assume the top- and bottom-heavy IMFs for calculating
the IMF-weighted yields of CCSNe. The circles illustrate the
metal enrichment scenario where the massive progenitors
($> 20 M_\odot$) undergo the HN channel \cite{Nomoto13} instead of CCSN.
The grey shaded area shows the SN~Ia fraction measured in the Perseus galaxy cluster core \cite{Simionescu19}.
}
\label{fig:snmodels}
\end{figure}

\clearpage
\renewcommand{\arraystretch}{1.3}

\begin{table}[htbp]
  \centering
  \caption{{\bf Energies of major fluorescence lines in the literature.}}\label{table:line_energy}
  \begin{tabularx}{\linewidth}{lllllll}
    \hline\hline
    Transition & Ref.\cite{Hoelzer97}$^{a}$ & Ref.\cite{Ito16}$^{b}$,\cite{Ito18}$^{c}$ & Ref.\cite{Bearden67} & xraylib$^{d}$ & NIST$^{e}$ & NIST$^{d,e}$\\
    & & & & (v4.1.5) & (experiment) & (theory) \\
    \hline
    S $K\alpha_1$ & ... & ... & 2307.84 & 2307.80 & 2307.885(34) & 2308.80(44)\\
    S $K\alpha_2$ & ... & ... & 2306.64 & 2306.60 & 2306.700(38) & 2307.01(45)\\
    S $K\beta$ & ... & ... & 2464.0 & 2464.00 & 2464.07(14) & 2469.73(60)\\

     &  &  & & 2464.00 & & 2467.53(72)\\

    Ar $K\alpha_1$ & ... & ... & 2957.70 & 2957.40 & 2957.682(16) & 2957.90(42)\\
    Ar $K\alpha_2$ & ... & ... & 2955.63 & 2955.30 & 2955.566(16) & 2955.89(43)\\
    Ar $K\beta$ & ... & ... & 3190.5 & 3190.10 & 3190.49(24) & 3191.31(58)\\
    &  &  & & 3190.20 & & 3191.47(58)\\

    Ca $K\alpha_1$ & ... & 3691.631(35) & 3691.68 & 3691.70 & 3691.719(49) & 3690.98(41)\\
    Ca $K\alpha_2$ & ... & 3688.105(37) & 3688.09 & 3688.10 & 3688.128(49) & 3687.56(43)\\
    Ca $K\beta$ & ... & 4012.644 & 4012.7 & 4012.70 & 4012.76(38) & 4014.32(59)\\

    & & & & 4012.70 & & 4014.68(58)\\

    Cr $K\alpha_1$ & 5414.81(1) & 5414.759(26) & 5414.72 & 5414.70 & 5414.8045(71) & 5413.88(42)\\
    Cr $K\alpha_2$ & 5405.54(1) & 5405.417(36) & 5405.51 & 5405.50 & 5405.5384(71) & 5404.06(45)\\
    Cr $K\beta$ & 5946.82(1) & 5946.758(42) & 5946.71 & 5946.70 & 5946.823(11) & 5940.74(92)\\
    & & 5945.556(46) & & 5946.70 & & 5947.10(100)\\

    Mn $K\alpha_1$ & 5898.80(1) & 5898.841(62) & 5898.75 & 5898.70 & 5898.8010(84) & 5898.10(42)\\
    Mn $K\alpha_2$ & 5887.59(1) & 5887.702(65) & 5887.65 & 5887.60 & 5887.6859(84) & 5886.20(45)\\
    Mn $K\beta$ & 6490.18(1) & 6490.57(11) & 5427.29 & 6490.40 & 6490.585(14) & 6485.39(96)\\

    & & 6489.17(13) & & 6490.40 & & 6492.7(10)\\

    Fe $K\alpha_1$ & 6404.01(1) & 6403.859(71) & 6403.84 & 6403.90 & 6404.0062(99) & 6403.13(43)\\
    Fe $K\alpha_2$ & 6391.03(1) & 6390.765(74) & 6390.84 & 6390.90 & 6391.0264(99) & 6389.51(46)\\
    Fe $K\beta$ & 7058.18(3) & 7059.989(66) & 7057.98 & 7058.00 & 7058.175(16) & 7053.23(100)\\
    & & 7059.989(66) & & 7058.00 & & 7059.9(11)\\

    Ni $K\alpha_1$ & 7478.26(1) & 7478.190(66) & 7478.15 & 7478.10 & 7478.2521(45) & 7477.72(44)\\
    Ni $K\alpha_2$ & 7461.04(1) & 7460.751(86) & 7460.89 & 7460.90 & 7461.0343(45) & 7459.96(47)\\
    Ni $K\beta$ & 8264.78(1) & 8264.98(14) & 8264.66 & 8264.70& 8264.775(17) & 8262.4(11)\\
    & & 8262.90(16) &  & 8264.70 &  & 8267.6(11)\\
    \hline
  \end{tabularx}
  \begin{flushleft}
  $^{a}$ Peak energies of the measured lines were used.\\
  $^{b}$ Fitting parameters from an asymmetric Lorentzian are used for $K\alpha_1$ and $K\alpha_2$.\\
  $^{c}$ For the $K\beta$ lines of Cr, Mn, Fe, and Ni, the values correspond to the $K\beta_1$ (first row) and $K\beta_3$ (second row) transitions.\\
  $^{d}$ For the $K\beta$ lines, the values correspond to the KM2 (first row) and KM3 (second row) transitions.\\
  $^{e}$ The data in this table were extracted from the X-ray Transition Energy Database at the National Institute of Standards and Technology (NIST) on March 19, 2025.\\\\
  Uncertainties are given in parentheses.\\
  \end{flushleft}
\end{table}

\end{document}